\documentclass[AMA,STIX1COL,doublespace]{WileyNJD-v2}

\articletype{RESEARCH ARTICLE}%

\received{26 April 2016}
\revised{6 June 2016}
\accepted{6 June 2016}

\begin{document}

\title{Statistical method for pooling categorical biomarkers from multi-center matched/nested case-control studies}

\author[1]{Yujie Wu}

\author[2]{Xiao Wu}

\author[3]{Mitchell H. Gail}

\author[3]{Regina G. Ziegler}

\author[4,5]{Stephanie A. Smith-Warner}

\author[1,4,6]{Molin Wang*}

\authormark{Wu \textsc{et al}}

\address[1]{\orgdiv{Department of Biostatistics}, \orgname{ Harvard T. H. Chan School of Public Health}, \orgaddress{\city{Boston}, \state{MA}}}

\address[2]{\orgdiv{Department of Biostatistics}, \orgname{ Columbia Mailman School of Public Health}, \orgaddress{\city{New York}, \state{NY}}}

\address[3]{\orgdiv{Division of Cancer Epidemiology and Genetics, National Cancer Institute}, \orgname{National Institutes of Health }, \orgaddress{\city{Bethesda}, \state{MD}}}

\address[4]{\orgdiv{Department of Epidemiology}, \orgname{ Harvard T. H. Chan School of Public Health}, \orgaddress{\city{Boston}, \state{MA}}}

\address[5]{\orgdiv{Department of Nutrition}, \orgname{ Harvard T. H. Chan School of Public Health}, \orgaddress{\city{Boston}, \state{MA}}}

\address[6]{\orgdiv{Channing Division of Network Medicine,  Department of Medicine, Brigham and Women's Hospital}, \orgname{ Harvard Medical School}, \orgaddress{\city{Boston}, \state{MA}}}

\corres{*Molin Wang, Departments of Biostatistics and Epidemiology, Harvard T. H. Chan School of Public Health, and Channing Division of Network Medicine, Department of Medicine, Brigham and Women's Hospital, Harvard Medical School, Boston, MA 02115.\\	\email{stmow@channing.harvard.edu}}

\abstract[Summary]{		Pooled analyses that aggregate data from multiple studies are becoming increasingly common in collaborative epidemiologic research in order to increase the size and diversity of the study population. However, biomarker measurements from different studies are subject to systematic measurement errors and directly pooling them for analyses may lead to biased estimates of the regression parameters. Therefore, study-specific calibration processes must be incorporated in the statistical analyses to address between-study/assay/laboratory variability in the biomarker measurements. We propose a likelihood-based method to evaluate biomarker-disease relationships for categorical biomarkers in matched/nested case-control studies. To account for the additional uncertainties from the calibration processes, we propose a sandwich variance estimator to obtain valid asymptotic variances of the estimated regression parameters. Extensive simulation studies with varying sample sizes and biomarker-disease associations are used to evaluate the finite sample performance of our proposed methods. As an illustration, we apply the methods to a vitamin D pooling project of colorectal cancer to evaluate the effect of categorical vitamin D levels on colorectal cancer risks.}

\keywords{Calibration, Conditional Likelihood, Matched Case-control Study, Measurement Error, Nested Case-control Study, Pooling Project}

\jnlcitation{\cname{%
		\author{<aurhor name>},
		\author{<aurhor name>},
		\author{<aurhor name>},
		\author{<aurhor name>},  and
		\author{<aurhor name>}} (\cyear{<year>}),
	\ctitle{<journal title>}, \cjournal{<journal name>} <year> <vol> Page <xxx>-<xxx>}

\maketitle

\section{Introduction}
\label{sec1}

Studies aiming to investigate biomarker-disease relationships often suffer from small sample sizes, and one way to increase sample sizes, as well as diversity in the study population, is to combine data from multiple studies. Pooling data from multiple studies can also enable potential examination of population subgroups and disease subtypes due to increased sample sizes \cite{key2010pooling, smith2006methods, tworoger2006use}.

There are two common approaches to analyzing  combined participant-level data from multiple studies: the two-stage method and the aggregated data method. The first stage of the two-stage method obtains study-specific exposure-disease associations, and in the second stage, a pooled estimate of the association is calculated by using meta-analytic techniques \cite{smith2006methods}. On the other hand, the aggregated data method performs a one-stage analysis on the pooled dataset that includes observations from all individual studies. One drawback of the two-stage method is that, in subset analyses or under rare outcome situations, data may become so sparse that some individual studies have no cases, or no controls, in some exposure categories, making the estimation of some coefficients difficult \cite{sloan2019design}. In contrast, the aggregated data method is less likely to encounter these data sparsity issues and allows more modelling flexibility \cite{kontopantelis2018comparison}. In this paper, we focus on developing methods for the aggregated data analysis {\color{black}{using participant-level data.}}

A major challenge in the analysis of pooled biomarker data is potential systematic differences in biomarker measurements among studies using different laboratories. Ignoring this between-assay/laboratory variability may lead to biased association estimates \cite{sloan2019design, sloan2021statistical}. For example, the Circulating Biomarkers and Breast and Colorectal Cancer Consortium (BBC3) was established to assess the associations between vitamin D, as measured by circulating concentrations of 25-hydroxyvitamin D [25(OH)D], the accepted measure of vitamin D status, and breast and colorectal cancer risk. 
The BBC3 consists of a total of 17 breast cancer nested case-control studies and 17 CRC nested case-control studies with 25(OH)D concentrations measured in different local laboratories, often using different assay methodologies, and the 25(OH)D measurements can vary up to 40\% across different laboratories and assays \cite{mccullough2018circulating}.

To address between-study variability in the biomarker measurements across studies, a study-specific calibration process needs to be incorporated.  This calibration process often relies on a designated reference laboratory, which uses a widely accepted, valid assay method.  Representative  subsets of the samples from each study are re-assayed in the reference laboratory, and we refer to these subsets as calibration subsets. In these calibration subsets, we can fit study-specific calibration models relating the reference laboratory measurements to the local laboratory measurements. Based on these calibration models, the reference laboratory measurements can be imputed for all study participants. Ideally, the calibration study samples should be selected at random from the entire study cohort. {\color{black}{However, due to potential concerns about the availability of biomarker measurements in cases, investigators typically only select controls into the calibration studies \cite{sloan2021statistical}}}. 

In our motivating BBC3 study, researchers were particularly interested in the relationship between breast and colorectal cancer risk and circulating 25(OH)D which were categorized based on the Institute of Medicine (IOM), now named the National Academy of Medicine, recommended guidance: deficiency ($<30$ nmol/L), insufficiency ($[30,50)$ nmol/L), sufficiency and beyond sufficiency ($\ge 50$ nmol/L)\cite{mccullough2018circulating}. Circulating 25(OH)D measurements were obtained from a variety of local laboratories in the studies and thus their concentrations may not be directly comparable on an absolute scale. Therefore, 29 controls from each nested case-control study were sampled to have their blood samples re-assayed in the reference laboratory. Sloan et al. (2019) presents methods for pooling biomarker data when the biomarker is treated as a continuous variable in statistical analyses \cite{sloan2021statistical}. {\color{black}{Here, we consider developing methods to pool categorical biomarkers. }}{\color{black}{The usual analysis of categorical biomarker measurements proceeds by first fitting a regression calibration model to calibrate the continuous local laboratory measurements, and then discretizing the calibrated continuous biomarker measurements for categorical analysis \cite{gail2016calibration}. In this paper, we consider an alternative approach that directly obtains the calibrated categorical biomarkers through a multinomial logistic regression, and we propose a joint likelihood approach for inferences on the association between categorical biomarker measurements and disease risks}}.  The proposed methods eliminate the bias in the effect estimates due to between-study/assay/laboratory variability in the biomarker measurements, and account for the additional uncertainties introduced by the calibration processes for inference.

 The paper is organized as follows. In Section 2, we introduce the methods; in Section 3, we perform simulation studies to evaluate our methods in finite sample sizes; Section 4 presents the real data application, evaluating the effect of categorical circulating 25(OH)D concentrations on colorectal cancer risk, and Section 5 concludes the paper. 


	\section{Methods}
	\label{sec2}
	
	\subsection{Notations and Models}
	Let $Y$ denote the binary disease outcome, $X$ denote the continuous biomarker measurement from the  reference laboratory, $W$ denote the continuous biomarker measurement from study-specific local laboratories, and $\mathbf{Z}$ be a column vector of other covariates. Assume there are $S$ matched/nested case-control studies, and  the $s$-th  study consists of $N_s$ strata, with one case and $M_{sj}$ matched controls in the $j$-th stratum, $j = 1,\ldots,N_s$, $s = 1,\ldots,S$. For notational simplicity, we assume there is only
	one case in each stratum. The method can be straightforwardly extended to designs with multiple cases in some or all strata. 
	
	We assume the following conditional logistic regression model for the biomarker-disease relationship:
	\begin{equation*}
		\textrm{logit}(P(Y_{sji} = 1)) = \beta_{0sj} + \boldsymbol{\beta}_X^T \boldsymbol{B}( X_{sji}) + \boldsymbol{\beta}_{\boldsymbol{Z}}^T \boldsymbol{Z}_{sji},
	\end{equation*}
	where  $\boldsymbol{B}(X_{sji})$ is a $(P-1)$-dimensional column vector of indicators corresponding to the $P$-level categorical version of $X$ for the $i$-th individual of the $j$-th stratum in study $s$; $\beta_{0sj}$ is the stratum-specific intercept, and of interest is $\boldsymbol{\beta}_X$, which, for nested case-control studies under  incidence density sampling,  represents the log relative risk (RR) of the biomarker measurements on the diseases \cite{breslow1978estimation, prentice1978retrospective}. We use superscript $T$ to denote the transpose of a vector or matrix throughout the paper.

	The biomarker exposures are measured in the study-specific `local' laboratories; that is, $(Y, W, \boldsymbol{Z})$ are available for all study participants; while the reference laboratory measurements, $X$, are only available for  individuals in the calibration subsets where a random blood sample from controls or both cases and controls were selected from each local laboratory to be re-measured in the reference laboratory. Thus $(Y, W, X, \boldsymbol{Z})$ are observed for participants in the calibration subsets.

	\subsection{Conditional likelihood with local laboratory measurements} \label{mle}
	
	If $X$ were known, we could estimate $\boldsymbol{\beta}=(\boldsymbol{\beta}_X,\boldsymbol{\beta}_{\boldsymbol{Z}})$ by maximizing the  likelihood function for conditional logistic regression. 
	The conditional likelihood function from the $j$-th stratum of the $s$-th study with a $1:M_{sj}$ matched case-control pair when the reference laboratory measurements are available is:
	\begin{equation}  
		\begin{split}
			L^\ast_{sj}\left(\boldsymbol{\beta};X,Y,\boldsymbol{Z}\right) &=\text{P}\left(Y_{sj1}=1,Y_{sj2}=0,\ldots, Y_{sj,M_{sj}+1}=0\mid \boldsymbol{X}_{sj}, \boldsymbol{Z}_{sj},\sum_{i=1}^{M_{sj}+1}Y_{sji}=1 \right)\\
			&=  \frac{\exp\left(\boldsymbol{\beta}_X^T \boldsymbol{B}\left( X_{sj1}\right)  + \boldsymbol{\beta_{\boldsymbol{Z}}}^T \boldsymbol{Z}_{sj1}\right) 
			}{\exp\left(\boldsymbol{\beta}_X^T \boldsymbol{B}\left( X_{sj1}\right)  + \boldsymbol{\beta_{\boldsymbol{Z}}}^T \boldsymbol{Z}_{sj1}\right) 
				+\sum^{M_{sj}+1}_{i=2}  \exp\left(\boldsymbol{\beta}_X^T \boldsymbol{B}\left( X_{sji}\right)  + \boldsymbol{\beta_{\boldsymbol{Z}}}^T \boldsymbol{Z}_{sji}\right), 
			}
		\end{split}
		\label{like_X}
	\end{equation}
	where $\boldsymbol{Z}_{sj}$ and $\boldsymbol{B}(\boldsymbol{X}_{sj})$ are  matrices containing the measurements of all participants from the $j$-th stratum of the $s$-th study. For notational simplicity, we use $i=1$ to index the case, and $i=2,\ldots,M_{sj}+1$ to index the matched controls.
	
	However, $L^\ast_{sj}$ cannot be directly calculated since the reference laboratory measurements $X$ are not available to all study participants; instead, most participants only have local laboratory measurements $W$. To derive the likelihood contribution based on the local laboratory measurements, we make the following  surrogacy assumption \cite{armstrong1985measurement,carroll2006measurement}:
	\begin{equation*}
		\text{P}\left(\boldsymbol{Y}_{sj}\mid \boldsymbol{X}_{sj}, \boldsymbol{Z}_{sj},\sum_{i=1}^{M_{sj}+1}Y_{sji}=1 \right)=	\text{P}\left(\boldsymbol{Y}_{sj}\mid \boldsymbol{X}_{sj},\boldsymbol{W}_{sj}, \boldsymbol{Z}_{sj},\sum_{i=1}^{M_{sj}+1}Y_{sji}=1 \right),
	\end{equation*}
	where $\boldsymbol{Y}_{sj}$ is the vector containing the outcome information for the $j$-th stratum of the $s$-th study. The surrogacy assumption implies that given the reference laboratory measurements, other covariates of interest and the study design, the local laboratory measurements will not provide additional information for the outcomes \cite{sloan2021statistical}.
	
	Under the surrogacy assumption, the likelihood contribution from a stratum using local laboratory measurements can be written as:
	\begin{equation} 
		\footnotesize
		\begin{split}
			\label{eq1}
			L_{sj}(\boldsymbol{\beta};W,Y,\boldsymbol{Z})&=\text{P}\left(\boldsymbol{Y}_{sj}\mid  \boldsymbol{W}_{sj}, \boldsymbol{Z}_{sj},\sum_{i=1}^{M_{sj}+1}Y_{sji}=1 \right)\\
			&=\int_{\boldsymbol{X}_{sj}}\text{P}\left(\boldsymbol{Y}_{sj}\mid  \boldsymbol{W}_{sj}, \boldsymbol{X}_{sj}, \boldsymbol{Z}_{sj},\sum_{i=1}^{M_{sj}+1}Y_{sji}=1 \right) \text{f}\left(\boldsymbol{X}_{sj}\mid  \boldsymbol{W}_{sj}, \boldsymbol{Z}_{sj},\sum_{i=1}^{M_{sj}+1}Y_{sji}=1 \right)\,d\boldsymbol{X}_{sj}\\
			&=\sum_{\boldsymbol{B}(\boldsymbol{X}_{sj})}\text{P}\left(\boldsymbol{Y}_{sj}\mid \boldsymbol{X}_{sj}, \boldsymbol{Z}_{sj},\sum_{i=1}^{M_{sj}+1}Y_{sji}=1 \right) \text{P}\left(  \boldsymbol{X}_{sj}\in\boldsymbol{R}_{sj}  \mid  \boldsymbol{W}_{sj}, \boldsymbol{Z}_{sj},\sum_{i=1}^{M_{sj}+1}Y_{sji}=1 \right)\\
			&= \sum^P_{p_{sj,M_{sj}+1}=1} \ldots\sum^P_{p_{sj2}=1}  \sum^P_{p_{sj1}=1} \Bigg[ P ( X_{sj1}\in R_{p_{sj1}} | W_{sj1}, \boldsymbol{Z}_{sj1},\sum_{i=1}^{M_{sj}+1}Y_{sji}=1) \prod^{M_{sj}+1}_{i=2} P ( X_{sji} \in R_{p_{sji}} | W_{sji}, \boldsymbol{Z}_{sji},\sum_{i=1}^{M_{sj}+1}Y_{sji}=1) \\
			&\times \frac{\exp\{\boldsymbol{\beta}_{X}^T \boldsymbol{B}(X_{sj1} \in R_{p_{sj1}})+ \boldsymbol{\beta_Z}^T \boldsymbol{Z}_{sj1}\} 
			}{\exp\{\boldsymbol{\beta}_{X}^T \boldsymbol{B}(X_{sj1} \in R_{p_{sj1}}) +\boldsymbol{\beta_{\boldsymbol{Z}}}^T \boldsymbol{Z}_{sj1}\}
				+\sum^{M_{sj}+1}_{i=2}  \exp\{\boldsymbol{\beta}_{X}^T \boldsymbol{B}(X_{sji}\in R_{p_{sji}}) + \boldsymbol{\beta_Z}^T \boldsymbol{Z}_{sji}\} 
		}\Bigg].
	\end{split}
\end{equation}
where $R_p$ is the range of the biomarker value corresponding to the $p$-th category,  $p=1,\ldots,P$.

The probability $P ( X_{sji} \in R_{p_{sji}} | W_{sji}, \boldsymbol{Z}_{sji},\sum_{i=1}^{M_{sj}+1}Y_{sji}=1)$ can be estimated from the calibration study subset. We make the calibration assumption that \cite{sloan2021statistical}:
\begin{equation*}
P ( X_{sji} \in R_{p_{sji}} | W_{sji}, \boldsymbol{Z}_{sji},\sum_{i=1}^{M_{sj}+1}Y_{sji}=1)\approx P ( X_{sji} \in R_{p_{sji}} | W_{sji},\sum_{i=1}^{M_{sj}+1}Y_{sji}=1),
\end{equation*}
where given the local laboratory measurements $W_{sji}$, other covariates of interest $\boldsymbol{Z}_{sji}$ do not provide any additional information on the reference laboratory measurements \cite{sloan2021statistical}. We model the underlying true probabilities through multinomial logistic regression for the $s$-th study \cite{agresti2012categorical}, in which, 
\begin{equation}
\log \frac{Pr(X_{sji}\in R_{p}|W_{sji},\sum_{i=1}^{M_{sj}+1}Y_{sji}=1;\boldsymbol{a}^{(s)}, \boldsymbol{b}^{(s)})}{Pr(X_{sji}\in R_{1}|W_{sji},\sum_{i=1}^{M_{sj}+1}Y_{sji}=1;\boldsymbol{a}^{(s)}, \boldsymbol{b}^{(s)})} = a_{p}^{(s)}+b_{p}^{(s)}W_{sji}, \quad\text{for } p=2,\ldots, P.
\label{logodds}
\end{equation}
The corresponding probabilities can therefore be estimated by,

\begin{equation} 
\begin{split}
	\label{eq2}
	\hat{P}(X_{sji}  \in R_{p} | W_{sji},\sum_{i=1}^{M_{sj}+1}Y_{sji}=1) &= \frac{1}{1+ \sum^P_{p'=2} \exp(\hat{a}^{(s)}_{p'}+\hat{b}^{(s)}_{p'} W_{sji}) } \ \ \ \ \textrm{for} \ p=1 \\
	\hat{P}(X_{sji}  \in R_{p} | W_{sji}, \sum_{i=1}^{M_{sj}+1}Y_{sji}=1) &= \frac{\exp(\hat{a}^{(s)}_{p}+\hat{b}^{(s)}_{p} W_{sji})}{1+\sum^P_{p'=2}  \exp(\hat{a}^{(s)}_{p'}+\hat{b}^{(s)}_{p'} W_{sji}) } \ \ \ \ \textrm{for} \ p>1,
\end{split}
\end{equation} 
where $\hat{a}^{(s)}_{p}$ and $\hat{b}^{(s)}_{p}, p=2,\ldots,P$ are estimated using data from the calibration subsets. Note that $a^{(s)}_{p}$ and $b^{(s)}_{p}$ are the study-specific intercept and slope for the $p$-th category in the multinomial logistic regression, and need to be estimated separately in each study. Plugging in the corresponding estimated probabilities, the {\color{black}{pseudo-likelihood}} contribution from the $j$-th stratum in study $s$ based on the local laboratory measurements is:
\begin{equation} 
\begin{split}
	\hat{L}_{sj} (W | Y,\boldsymbol{\beta} ) 		 &= \sum^P_{p_{sj,M_{sj}+1}=1} \ldots\sum^P_{p_{sj2}=1}  \sum^P_{p_{sj1}=1} \Bigg[ \widehat{P}( X_{sj1}\in R_{p_{sj1}} | W_{sj1}, \sum_{i=1}^{M_{sj}+1}Y_{sji}=1) \prod^{M_{sj}+1}_{i=2} \widehat{P} ( X_{sji} \in R_{p_{sji}} | W_{sji},\sum_{i=1}^{M_{sj}+1}Y_{sji}=1) \\
	&\times \frac{\exp\{\boldsymbol{\beta}_{X}^T B(X_{sj1} \in R_{p_{sj1}})+ \boldsymbol{\beta_Z}^T \boldsymbol{Z}_{sj1}\} 
	}{\exp\{\boldsymbol{\beta}_{X}^T B(X_{sj1} \in R_{p_{sj1}}) +\boldsymbol{\beta_Z}^T \boldsymbol{Z}_{sj1}\}
		+\sum^{M_{sj}+1}_{i=2}  \exp\{\boldsymbol{\beta}_{X}^T B(X_{sji}\in R_{p_{sji}}) + \boldsymbol{\beta_Z}^T \boldsymbol{Z}_{sji}\} 
}\Bigg].
\end{split}
\label{cond_likelihood}
\end{equation}

{\color{black}{Due to concerns of the availability of biomarker measurements in cases,}} investigators typically select only controls into the calibration subsets \cite{gail2016calibration}. This leads to the statistical challenge that the real calibration model fitted in the $s$-th study is:

\begin{equation*}
\log \frac{Pr(X_{sji}\in R_{p}|W_{sji},Y_{sji}=0;\boldsymbol{a}_{co}^{(s)}, \boldsymbol{b}_{co}^{(s)})}{Pr(X_{sji}\in R_{1}|W_{sji},Y_{sji}=0;\boldsymbol{a}_{co}^{(s)}, \boldsymbol{b}_{co}^{(s)})} = a_{p,co}^{(s)}+b_{p,co}^{(s)}W_{sji}, \quad\text{for } p=2,\ldots, P.
\end{equation*}
In general, $\widehat{{a}}^{(s)}_{p,co}$ and $\widehat{{b}}^{(s)}_{p,co}$ are not consistent estimates of ${{a}}^{(s)}_{p}$ and ${{b}}^{(s)}_{p}$. We present the theoretical results in Appendix~\ref{appendix_cal} showing that if the effect size is small, that is $\boldsymbol{\beta}_X \approx \boldsymbol{0}$, then ${P}(X_{sji}  \in R_{p} | W_{sji},\sum_{i=1}^{M_{sj}+1}Y_{sji}=1)\approx {P}(X_{sji}  \in R_{p} | W_{sji},Y_{sji}=0)$.

Since participants in the calibration subsets have both reference and local laboratory measurements, following Sloan et al. (2021) \cite{sloan2021statistical}, we propose the Full calibration and Internalized calibration methods. For the Full calibration method, the local biomarker measurements are always calibrated no matter whether the participant has a reference laboratory measurement available and therefore the parameters are estimated by directly solving Equation (\ref{cond_likelihood}) across strata and studies. For the Internalized calibration method, the calibration processes will only be applied to individuals outside the calibration subsets; and therefore for participants in the calibration subsets, their likelihood contribution in Equation (\ref{cond_likelihood}) has the corresponding probabilities: ${P}(X_{sji}  \in R_{p} | W_{sji},\sum_{i=1}^{M_{sj}+1}Y_{sji}=1)=1$ if the observed biomarker measurements from the reference laboratory falls into the $p$-th category and 0 otherwise. 

\subsection{Parameter estimates and inference}

Let $\boldsymbol{a}=[a_2^{(1)}, a_3^{(1)},\ldots,a_P^{(1)},a_2^{(2)}, a_3^{(2)},\ldots,a_P^{(2)},\ldots,a_2^{(S)}, a_3^{(S)},\ldots,a_P^{(S)}]^T$ and \\$\boldsymbol{b}=[b_2^{(1)}, b_3^{(1)},\ldots,b_P^{(1)},b_2^{(2)}, b_3^{(2)},\ldots,b_P^{(2)},\ldots,b_2^{(S)}, b_3^{(S)},\ldots,b_P^{(S)}]^T$ be the collection of the parameters associated with the study-specific multinomial logistic regressions fitted in the calibration subsets. The collection of parameters that need to be estimated are therefore $\boldsymbol{\theta}=[\boldsymbol{a}^T,\boldsymbol{b}^T,\boldsymbol{\beta}^T]^T$, and the corresponding joint estimating equations are $
[\boldsymbol{\psi}_{\boldsymbol{a}}^T, \boldsymbol{\psi}_{\boldsymbol{b}}^T,
\boldsymbol{\psi}_{\boldsymbol{\beta}_X}^T,   
\boldsymbol{\psi}_{\boldsymbol{\beta}_{\boldsymbol{Z}}}^T  ]^T=\boldsymbol{0}$. We provide the technical details in the supplementary material. 

In practice, we adopt a two-step pseudo maximum likelihood approach to obtain the parameter estimates \cite{Gong1981pseudo}. In the first step, we fit the multinomial logistic regressions in the calibration subsets and obtain estimates for ${\boldsymbol{a}}$ and ${\boldsymbol{b}}$; and in the second step, we solve $[ \boldsymbol{\psi}_{\boldsymbol{\beta}_X}(\widehat{\boldsymbol{a}}, \widehat{\boldsymbol{b}})^T ,  \boldsymbol{\psi}_{\boldsymbol{\beta}_{\boldsymbol{Z}}}(\widehat{\boldsymbol{a}}, \widehat{\boldsymbol{b}})^T    ]^T =\boldsymbol{0}$ to get the estimated regression parameters $\widehat{\boldsymbol{\beta}}_X$ and $\widehat{\boldsymbol{\beta}}_Z$, where we plug in the estimated calibration model parameters $\widehat{\boldsymbol{a}}$ and $\widehat{\boldsymbol{b}}$ into the estimating equations for ${\boldsymbol{\beta}}_X$ and ${\boldsymbol{\beta}}_Z$.

To account for the uncertainty in the estimation of the calibration model parameters, we propose to use the sandwich variance estimator by stacking all estimating equations together to get the robust estimates of the regression parameters $\widehat{\boldsymbol{\theta}}$; detailed formulae are provided in the supplementary material. 

\section{Simulation studies}
\label{sec3}

We conduct simulation studies to evaluate the finite-sample performance of our proposed methods using both the Full calibration (FC) and Internalized calibration (IN) methods. Percent biases and empirical coverage rates of the 95\% confidence intervals are reported. 


{\color{black}{The parameters in the simulations are based in part upon the scenarios considered in Gail et al. (2016) \cite{gail2016calibration} and Sloan et al. (2021) \cite{sloan2021statistical}}}.  We generate $S=3$ independent studies, containing $N_s=500$ matched $1:1$ case-control pairs with local laboratory measurements. For each observation, $W_{sji}$ is generated from a normal distribution with standard deviation $16$ and mean 33.8657, 41.3204, 47.7603, respectively, for $s=1, 2, 3$. We first consider the simple situation where a three-level categorical reference laboratory measurement is directly generated from a multinomial distribution, with the following log odds:
\begin{equation*}
\begin{split}
\log \frac{Pr(X_{sji}\in R_{2}|W_{sji},\sum_{i=1}^{M_{sj}+1}Y_{sji}=1;\boldsymbol{a}^{(s)}, \boldsymbol{b}^{(s)})}{Pr(X_{sji}\in R_{1}|W_{sji},\sum_{i=1}^{M_{sj}+1}Y_{sji}=1;\boldsymbol{a}^{(s)}, \boldsymbol{b}^{(s)})} &= a_2^{(s)}+b_2^{(s)}W_{sji}\\
\log \frac{Pr(X_{sji}\in
	R_{3}|W_{sji},\sum_{i=1}^{M_{sj}+1}Y_{sji}=1;\boldsymbol{a}^{(s)}, \boldsymbol{b}^{(s)})}{Pr(X_{sji}\in R_{1}|W_{sji},\sum_{i=1}^{M_{sj}+1}Y_{sji}=1;\boldsymbol{a}^{(s)}, \boldsymbol{b}^{(s)})} &= a_3^{(s)}+b_3^{(s)}W_{sji}.\\
\end{split}
\end{equation*}
where $a_2^{(s)}\in\{-13.7925, -13.5167, -13.5219\}$, $b_2^{(s)}\in\{0.3259, 0.3203, 0.3246\}$, $a_3^{(s)}\in\{-29.8290, -29.6692, -29.7058\}$ and $b_3^{(s)}\in\{0.6324, 0.6348, 0.6427\}$.

The outcome model takes the following form:
\begin{equation}
\textrm{logit}(P(Y_{sji} = 1)) = \beta_{0sj} + \boldsymbol{\beta}_X^T \boldsymbol{B}( X_{sji}),
\label{sim_outcome}
\end{equation}
where $\beta_{0sj}$ is generated from $N(-1,0.1^2)$, and we set the effect sizes of category 2 and 3 with respect to category 1 (the reference group) to be $[-\log(1.25)/2,-\log(1.25) ]$ or $[-\log(2)/,-\log(2) ]$, respectively. For each study, we randomly select $n_v=50$ or $100$ controls into the calibration subsets that also have the categorical reference biomarker measurements available. 

Table ~\ref{cate_x_1} reports the percent biases and empirical coverage rates of the 95\% confidence intervals for both the full and internalized calibration methods. The mean of the estimated standard errors (SE) using the sandwich variance estimator and the empirical standard error (ESE) of the estimated coefficients over the 1000 simulation replicates are also reported. The percent biases of the point estimates from the two calibration methods are generally less than 5\%, with the full calibration method being less biased. The coverage rates of the 95\% confidence intervals are centered around the 95\% nominal level, with the mean of the estimated standard errors of the regression parameters using the sandwich variance estimator being close to the empirical standard errors. Moreover, we observe less biased estimates when the effect size is small than when the effect size is large, which is consistent with our theoretical justification. 

 
{\color{black}{To compare our methods to the method that directly discretizes calibrated continuous biomarkers \cite{gail2016calibration},}} we additionally generate the continuous reference laboratory measurement $X$ first and categorize it based on some pre-specified cut points. We generate the local and reference laboratory measurements from the following bivariate normal distribution:

\begin{equation*}
(W_{sji}, X_{sji})\sim \text{MVN}\left(\begin{bmatrix}
\mu_{W_s}\\
a+b\mu_{W_s}
\end{bmatrix},  \begin{bmatrix}
\tau^2+\sigma_{WW}&b\tau^2+\sigma_{XW}\\
b\tau^2+\sigma_{XW}&b^2\tau^2+\sigma_{XX}
\end{bmatrix} \right)
\end{equation*}
where we set $\tau^2=240.89, \sigma_{WW}=\sigma_{XX}=16$, $a=5, b=1.4$ and $\mu_{W_s}=33.87, 41.32, 47.76$ for $s=1,2,3$. We then categorize the reference laboratory measurements $X$ based on the thresholds 62.9 and 76.3 into three levels, and the level with the lowest $X$ value is set to be the reference group. The data generation model for the outcome still follows Equation ~\ref{sim_outcome}. In simulation, for comparison, we also consider a naive approach, where no calibration is performed and we directly pool the local laboratory measurements together and categorize $W_{sji}$  based on the pre-specified thresholds. We additionally consider another measurement error calibration approach, where in the calibration subsets, we fit the linear regression models: $E(X_{sji}|W_{sji})=\gamma_s+\eta_sW_{sji}$, and get the imputed continuous reference laboratory measurements $\widehat{X}_{sji}$ for participants in each study. Then we create the categorical reference laboratory measurements by cutting the imputed reference laboratory measurements into three levels and fit the conditional logistic regression. We refer to this method as the linear calibration approach.

Table ~\ref{continuous_x_1} reports the simulation results. We observe substantially biased estimates of the parameters from the naive approach, where the percent biases are generally greater than 30\% with under-covered 95\% confidence intervals. The linear calibration approach is less biased than the naive approach. However, the percent biases are still greater than 10\% and the 95\% confidence intervals are also under covered. Our proposed methods have the lowest percent biases that are in general less than 5\%, and the coverage rates of the 95\% confidence intervals are close to the 95\% nominal level. In addition, the full calibration method is also less biased than the internalized calibration method. 

\section{Real data application}
We apply the proposed methods to two cohort studies participating in VDPP, the Nurses Health Study I (NHS) \cite{colditz1997nurses} and the Health Professionals Follow-up Study (HPFS) \cite{grobbee1990coffee}, to estimate the association between vitamin D, as measured by circulating 25(OH)D concentrations, and colorectal cancer (CRC) risk. Both studies are large prospective cohort studies in the United States. The HPFS enrolled 51,529 male health professionals in 1986 aged 40 to 75 years at baseline. The NHS began enrollment in 1976 and includes 121,701 female nurses aged 30 to 55 years at baseline. Between 1989 and 1997, each study obtained blood samples from a subset of participants. Circulating 25(OH)D was measured in a colorectal cancer nested case-control study in each cohort. The controls in each study were selected using incidence density sampling and were matched on sex, age, date of blood draw, and other study-specific matching factors in the two studies. {\color{black}{There was a total of 615 matched strata (267 from HPFS, 348 from NHS) and most strata included 2 controls per case (596 strata). }}Calibration subsets were obtained through stratified sampling so that there were 2-3 controls from each decile of the control distribution, and their blood samples were re-assayed at Heartland Assays, LLC (Ames, IA) during 2011-2013 \cite{mccullough2018circulating}. 


Our interest lies in estimating the relative risks (RRs) between CRC and different categories of circulating 25(OH)D, {\color{black}{where we categorize the circulating 25(OH)D concentrations based on IOM recommendations, i.e. deficiency ($<30$ nmol/L), insufficiency ($[30,50)$ nmol/L), sufficiency and beyond sufficiency ($\ge 50$ nmol/L) \cite{mccullough2018circulating}}}. We compare the estimated RR using 1) our proposed method for both the full calibration and internalized calibration methods, and 2) the naive method that directly categorizes the local laboratory measurements. The data application results are presented in Table~\ref{apptable}.  Both methods demonstrate that higher 25(OH)D levels are associated with reduced CRC risk. {\color{black}{Our proposed method yields more pronounced effect estimates compared to the naive method. Furthermore, the reduction in CRC risk at sufficient and beyond sufficient circulating 25(OH)D concentrations, compared to deficient concentrations, reached statistical significance using our proposed full or internalized calibration methods, but not with the naive method.}}

\section{Discussion}
\label{sec4}
In this paper, we develop methods to analyze pooled biomarker data from multiple nested/matched case-control studies by incorporating a study-specific calibration process to address the between-study/assay/laboratory variation in the biomarker measurements. A sandwich variance estimator is proposed to estimate the variance of the estimated regression coefficients to account for the additional uncertainty introduced by the calibration process. 

In simulations, we show that naively pooling circulating biomarker data with variation in absolute concentrations due to assay and laboratory measurement differences without calibration will result in substantial bias in the pooled estimate of the biomarker-disease association. In contrast, our methods, both the full calibration (FC) and internalized calibration (IN) methods obtain less biased estimates even with relatively small sample sizes with coverage rates of the 95\% confidence intervals based on our proposed sandwich variance estimator being close to the 95\% nominal level. The full calibration method performs slightly better than the internalized calibration method in terms of percent bias under all simulation scenarios. One reason for the smaller bias of the full calibration method compared to the internalized calibration method could be that {\color{black}{our proposed pseudo-likelihood contribution from the $j$-th stratum in study $s$ represents the following conditional probability $\text{P}\left(\boldsymbol{Y}_{sj}\mid  \boldsymbol{W}_{sj}, \boldsymbol{Z}_{sj},\sum_{i=1}^{M_{sj}+1}Y_{sji}=1 \right)$, and consistent estimates can be obtained by maximizing the likelihood conditional on local laboratory measurements. However, for the internalized calibration method, we are in fact using $P(Y_{sji}|X_{sji}, Z_{sji},\sum_{i=1}^{M_{sj}+1}Y_{sji}=1)$ in place of $P(Y_{sji}|W_{sji}, Z_{sji},\sum_{i=1}^{M_{sj}+1}Y_{sji}=1)$ for controls who have reference laboratory measurements, while for the cases in the same stratum, the likelihood is still conditional on local laboratory measurements. Therefore, for the internalized calibration method, the pseudo-likelihood contribution from a stratum whose control is also in the calibration subset no longer represents the conditional probability of the outcome given the local laboratory measurements in the likelihood function, and thus bias could be introduced. }}

The novel aggregated data method introduced in this paper is the first likelihood-based approach which allows for pooling categorical biomarker data from multiple nested/matched case-control studies. The R code for implementing the proposed methods can be found at the corresponding author's website: \url{https://www.hsph.harvard.edu/molin-wang/software/}.





\section*{Acknowledgments}
We are grateful to Tao Hou and Shiaw-Shyuan (Sherry) Yaun for their assistance in accessing the data. We also thank the Circulating Biomarkers and Breast and Colorectal Cancer Consortium (BBC3). We also thank the participating cohorts and researchers in the Circulating Biomarkers and Breast and Colorectal Cancer Consortium. The consortium, including calibration of previously measured studies, was supported by the National Cancer Institute, National Institutes of Health (R01 CA152071); the Intramural Research Program of the Division of Cancer Epidemiology and Genetics, National Cancer Institute, National Institutes of Health; and the Breast Cancer Research Foundation.  M.W. was supported by NIH grants R03-CA212799-01, and U01CA16755.
{\it Conflict of Interest}: None declared.

	\bibliography{Reference}

\begin{table}[h!]
	\centering
	\caption{The categorical reference laboratory measurements are generated directly from a multinomial distribution. Percent bias, empirical coverage rates (CR) of the 95\% confidence intervals, the mean estimated standard error (SE) from the sandwich variance estimator and the empirical standard error (ESE) of the estimated coefficients over the 1000 simulation replicates are reported. }
	\begin{tabular}{cccccc}
		\hline\hline
		Calib. size&$\beta_x^*$&Calibration mthod  & Percent bias (\%) &  CR & SE (ESE)\\\hline
				50&\multirow{2}{*}{$-\log(1.5)/2$}
		&   Internalized  &-1.8&0.965&0.140 (0.137)\\
		&& Full Calibration  &-0.4&0.958&0.145(0.148)\\\cmidrule{2-6}
		&\multirow{2}{*}{$-\log(1.5)$}
		&   Internalized  &-2.6&0.947&0.103 (0.100)\\
		&& Full Calibration  &-1.6&0.947&0.103 (0.102)\\\hline
		100&\multirow{2}{*}{$-\log(1.5)/2$}
		&   Internalized  &-3.3&0.979&0.140 (0.128)\\
		&& Full Calibration  &-0.4&0.965&0.150 (0.149)\\\cmidrule{2-6}
		&\multirow{2}{*}{$-\log(1.5)$}
		&   Internalized  &-2.1&0.956&0.103 (0.100)\\
		&& Full Calibration  &0.1&0.954&0.104 (0.102)\\\hline
				{50}&{$-\log(2)/2$}
		&   Internalized  &-6.9&0.956&0.142 (0.144)\\
		&& Full Calibration  &-5.6&0.943&0.147 (0.156)\\\cmidrule{2-6}
		&{$-\log(2)$}
		&   Internalized  &-2.2&0.941&0.107 (0.108)\\
		&& Full Calibration  &-1.2&0.941&0.108 (0.110)\\\hline
				{100}&{$-\log(2)/2$}
		&   Internalized  &-6.9&0.966&0.141 (0.130)\\
		&& Full Calibration  &-4.2&0.956&0.152 (0.150)\\\cmidrule{2-6}
		&{$-\log(2)$}
		&   Internalized  &-3.0&0.943&0.107 (0.106)\\
		&& Full Calibration  &-0.8&0.944&0.108 (0.109)\\
		\hline  \hline 
		\multicolumn{6}{l}{\textsuperscript{*}\footnotesize{For each simulation scenario, the two values of $\beta_x$ represent the regression coefficients of the second and third category of }}   \\
		\multicolumn{6}{l}{\footnotesize{the exposure compared to the reference level (the first category).    }}    
	\end{tabular} 
	\label{cate_x_1}
\end{table}

\begin{sidewaystable}[h!]
	\centering
	\caption{Simulation results where the continuous biomarker measurements are first generated and then categorized based on pre-specified cut points. Percent bias, empirical coverage rates (CR) of the 95\% confidence intervals, the mean estimated standard error (SE) from the sandwich variance estimator and the empirical standard error (ESE) of the estimated coefficients over the 1000 simulation replicates are reported. $\boldsymbol{\beta}_{New}$ denotes our proposed method, $\boldsymbol{\beta}_{Linear}$ denotes the additionally considered linear calibration approach and $\boldsymbol{\beta}_{Naive}$ denotes the naive approach. }
	\begin{tabular}{cccccccccccc}
		\hline\hline\multirow{2}{*}{Calib. size}&\multirow{2}{*}{$\beta_x^*$}&\multirow{2}{*}{Calibration method}   &  \multicolumn{3}{ c }{Percent bias (\%)}   &    \multicolumn{3}{ c }{CR} & \multicolumn{3}{ c }{SE (ESE)} \\
		&&  & $\beta_{New}$&$\beta_{Linear}$&$\beta_{Naive}$&$\beta_{New}$&$\beta_{Linear}$&$\beta_{Naive}$&$\beta_{New}$&$\beta_{Linear}$&$\beta_{Naive}$\\
		\cmidrule(l{2pt}r{2pt}){2-3}\cmidrule(l{2pt}r{2pt}){4-6}\cmidrule(l{2pt}r{2pt}){7-9}\cmidrule(l{2pt}r{2pt}){10-12}
				\multirow{4}{*}{50}&\multirow{2}{*}{$-\log(1.5)/2$}
		&   Internalized  &-5.4&-15.0&\multirow{2}{*}{28.6}&0.956&0.944&\multirow{2}{*}{0.945}&0.134 (0.139)&0.094 (0.095)&\multirow{2}{*}{0.142 (0.141)}\\		&& Full Calibration  &-4.2&-13.5&&0.951&0.950&&0.145 (0.144)&0.094 (0.095)&\\ \cmidrule(l{2pt}r{2pt}){2-12}		&\multirow{2}{*}{$-\log(1.5)$}
		&   Internalized  &-2.6&-11.1&\multirow{2}{*}{38.0}&0.951&0.921&\multirow{2}{*}{0.910}&0.103 (0.103)&0.093 (0.094)&\multirow{2}{*}{0.292 (0.286)} \\
		&& Full Calibration  &-1.5&-12.0&&0.950&0.912&&0.104 (0.103)&0.093 (0.094)&\\\hline
		\multirow{4}{*}{100}&\multirow{2}{*}{$-\log(1.5)/2$}
		&   Internalized  &-6.6&-16.0&\multirow{2}{*}{28.5}&0.971&0.942&\multirow{2}{*}{0.941}&0.138 (0.128)&0.095 (0.094)&\multirow{2}{*}{0.142 (0.143)}\\
		&& Full Calibration  &-4.2&-13.0&&0.961&0.945&&0.148 (0.149)&0.095 (0.094)&\\ \cmidrule(l{2pt}r{2pt}){2-12}
		&\multirow{2}{*}{$-\log(1.5)$}
		&   Internalized  &-3.0&-10.3&\multirow{2}{*}{-38.1}&0.960&0.929&\multirow{2}{*}{0.921}&0.102 (0.103)&0.093 (0.093)&\multirow{2}{*}{0.287 (0.292)}\\
		&& Full Calibration  &-0.8&-12.2&&0.961&0.916&&0.104 (0.106)&0.093 (0.094)&\\ \hline
				\multirow{4}{*}{50}&\multirow{2}{*}{$-\log(2)/2$}
		&   Internalized  &-8.6&-16.4&\multirow{2}{*}{31.5}&0.936&0.900&\multirow{2}{*}{0.888}&0.147 (0.141)&0.098 (0.095)&\multirow{2}{*}{0.152 (0.147)}\\
		&& Full Calibration  &-7.5&-15.3&&0.927&0.904&&0.159 (0.140)&0.098 (0.095)&\\ \cmidrule(l{2pt}r{2pt}){2-12}
		&\multirow{2}{*}{$-\log(2)$}
		&   Internalized  &-2.6&-11.8&\multirow{2}{*}{-34.8}&0.938&0.868&\multirow{2}{*}{0.866}&0.109 (0.107)&0.095 (0.096)&\multirow{2}{*}{0.302 (0.298)} \\
		&& Full Calibration  &-1.6&-12.3&&0.939&0.857&&0.110 (0.108)&0.096 (0.096)&\\\hline
				\multirow{4}{*}{100}&\multirow{2}{*}{$-\log(2)/2$}
		&   Internalized  &-10.1&-18.1&\multirow{2}{*}{31.6}&0.948&0.891&\multirow{2}{*}{0.896}&0.137 (0.140)&0.096 (0.096)&\multirow{2}{*}{0.147 (0.151)}\\
		&& Full Calibration  &-7.8&-16.0&&0.935&0.907&&0.150 (0.158)&0.095 (0.097)&\\ \cmidrule(l{2pt}r{2pt}){2-12}
		&\multirow{2}{*}{$-\log(2)$}
		&   Internalized  &-2.8&-11.2&\multirow{2}{*}{-34.8}&0.944&0.877&\multirow{2}{*}{0.868}&0.107 (0.107)&0.095 (0.096)&\multirow{2}{*}{0.298 (0.301)}\\
		&& Full Calibration  &-0.6&-12.2&&0.939&0.855&&0.108 (0.110)&0.096 (0.096)&\\ 
		\hline  \hline      
				\multicolumn{12}{l}{\textsuperscript{*}\footnotesize{For each simulation scenario, the two values of $\beta_x$ represent the regression coefficients of the second and third category of the exposure compared to the reference level (the first category).  }}       
	\end{tabular} 
	\label{continuous_x_1}
\end{sidewaystable}

\begin{table}
	\centering
	\caption{Data Application Results: Relative risks (RR) and 95\% confidence intervals of colorectal cancer by IOM-based categories of circulating 25(OH)D.} 
	\label{apptable}
	\begin{tabular}{cccc}
		\hline\hline
		Method &  \multicolumn{3}{c}{Circulating 25(OH)D, nmol/L} \\
		\hline
		&$<30$&$[30,50)$&$\ge 50$\\\hline
		Naive & 1 (reference)& 0.84 [0.48, 1.48] & 0.64 [0.37, 1.09] \\ 
		\hline
		Full calibration &1 (reference) &  0.65 [0.35, 1.23] &  0.54 [0.33, 0.89] \\\hline
		Internalized calibration&1 (reference)& 0.65 [0.40, 1.07] &  0.55 [0.36, 0.82] \\
		\hline\hline
	\end{tabular}
\end{table}
	\section{Appendix \label{appendix}}
	\subsection{Mathematical details for control-only calibration \label{appendix_cal}}
	The validity of control-only calibration requires the condition of a small or null effect of $X$ on $Y$. Under the surrogate assumption, the distribution of $P ( X_{case} | W_{case})$ and $P ( X_{ctrl} | W_{ctrl})$ can be rewritten as,
	\begin{align*}
		P ( X_{case}| W_{case}) = Pr(X|W,Y = 1) &= \frac{f(X, W,Y = 1)}{f(W,Y = 1)} \\
		&= \frac{Pr(Y=1|X)f(X,W)}{\sum_X Pr(Y=1|X)f(X,W)}  \\
		P ( X_{ctrl}  | W_{ctrl}) = Pr(X|W,Y = 0) &= \frac{f(X, W,Y = 0)}{f(W,Y = 0)} \\
		&= \frac{Pr(Y=0|X)f(X,W)}{\sum_X Pr(Y=0|X)f(X,W)} 
	\end{align*}
	Under small or null effect of X on Y, we have $Pr(Y=1|X)$ and $Pr(Y=0|X)$ are approximately constant $\forall X$, then,
	\begin{align*}
		P ( X_{case} | W_{case} ) \approx \frac{f(X,W)}{\sum_X f(X,W)} \approx P ( X_{ctrl} | W_{ctrl} )
	\end{align*}
	In another hand, by solving the following equations, $\forall X, W$
	\begin{align*}
		\frac{Pr(Y=1|X)f(X,W)}{\sum_X Pr(Y=1|X)f(X,W)}  &= \frac{Pr(Y=0|X)f(X,W)}{\sum_X Pr(Y=0|X)f(X,W)} \\
		Pr(Y=1|X) &= 1- Pr(Y=0|X)
	\end{align*}
	We obtain $Pr(Y=1|X)$ is constant, that is equivalent to null effect of $X$ on $Y$. Therefore, we prove $P ( X_{case}  | W_{case}) = P ( X_{ctrl} | W_{ctrl})$ iff $\boldsymbol{\beta_1} = \boldsymbol{0}$.


	
	

\end{document}